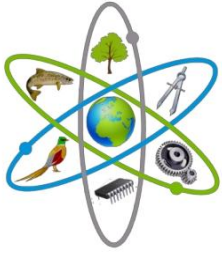

# Grayscale Image Authentication using Neural Hashing

Yakup Kutlu[1] and Apdullah Yayık[2*]

[1]Department of Computer Engineering, Faculty of Electrical and Electronics, İskenderun Technical University, Hatay, Turkey
[2]Turkish Army Forces

**Abstract**
Many different approaches for neural network based hash functions have been proposed. Statistical analysis must correlate security of them. This paper proposes novel neural hashing approach for gray scale image authentication. The suggested system is rapid, robust, useful and secure. Proposed hash function generates hash values using neural network one-way property and non-linear techniques. As a result security and performance analysis are performed and satisfying results are achieved. These features are dominant reasons for preferring against traditional ones.

**Introduction**

Cryptography uses mathematical techniques for information security. Information security is now a compulsory component of commercial applications, military communications and also social media implementations. So it can be said that cryptography is, furthermore, the most significant part of communication security (Arvandi et al. 2006). It maintains the condentiality that is the core of information security. Any cryptography requires condentiality, authentication, integrity and non-repudiation from those authorized to have it. Authentication relates to the identication of two parties entering into communication, while integrity addresses the unauthorized modification of an element inserted into the system (Sağıroğlu & Özkaya 2007). To date, there has been a large number of studies intended to advance robust cryptosystems and use them in communications. It is a novel and growing technique to perform the non-linear property of neural network to create secure hash function. Hash functions converts major definitions to minor values. As an input any message can be used, and as an output fixed length value is produced. Most popular hash functions are MD-5 (R. Rivest: 1992), SHA-2 that is by NSA (National Security Agency) and published in 2002 by the NIST as a U.S FIPS (Federal Information processing Standard) and SHA-3 that is based on an instance of the KECCAK algorithm that NIST selected as the winner of the Cryptographic Hash Algorithm Competition in 2013. Hash functions are used for data integrity and digital signature. Digital signature signs data in order to prove the accuracy of data and ID of sender. Hash function is digest of message which is attached to original message. Any modifying in original message makes hash function disabled. In other words; hash function is an information generating process from any message using mathematical techniques. Generated digital information is called message digest. Recycling of hash function must be almost impossible, so hash function must not inspire anyone about original message. It must be impossible to predict different messages whose hash values are the same. Hash value of every message is different, so any modifying in the original message makes digital signature invalid. Cryptography needs functions like this because they are able to provide safety communications (Soyalıç 2005). Hash functions are also used in Network and Internet Security. Any domain controlled PC client password is saved in _le server manager as its hash value, so administrator cannot do see clients original passwords. Also any malicious access to server database cannot capture client's original passwords. Up to now, there have been lots of studies to advance robust machine learning based hash-functions and use them in communications (Zou & Xiao 2009; Lian et al. 2007; YAYIK & KUTLU n.d.; Yayık & Kutlu 2013; Huang 2011). Near past and recently there is relatively much interest in using neural networks for cryptography (Lian et al. 2006) . Statistical analysis for sensitivity of SHA-2 secure hash algorithm and neural based hash function are nearly same (Sumangala et al. 2011), so it can be said that neural network will be used in cryptology in near future. There are many different approaches for image hash function algorithms. In 2011, Radon Transform based image fingerprinting (hashing) is proposed (Seo et al. 2004). Monga and Evans extracted vital image features using wavelet-based feature detection algorithm in order to advance image hashing system (Monga et al. 2006). In 2006, rotation invariance of Fourier-Mellin transform and controlled randomization based image hashing algorithm is introduced (Swaminathan et al. 2006). In last two decades neural network based hash function is studied by some researchers (Zou & Xiao 2009; Lian et al. 2007; YAYIK & KUTLU n.d.; Yayık & Kutlu 2013; Sumangala et al. 2011). Common feature of these neural network based related works is considering gray scale images or texts. In this paper, secure and robust neural image hash function for gray scale image, which uses non-linearity. Then, many experiments are performed to validate its security and statistical requirements. The rest of this paper is organized as follows. Section 2 describes proposed image hash function. Section 3 presents experimental results and performance analysis. Finally conclusion is given in Section 4.

---

[*] *Corresponding Author :Apdullah YAYIK, email: ayayik@kkk.tsk.tr*

**Materials and Methods**

*Proposed Model*

In the proposed hash function neural network shown in Figure 1. is used which has three layers that carries out ideal hash functions confusion, diffusion and compression properties.

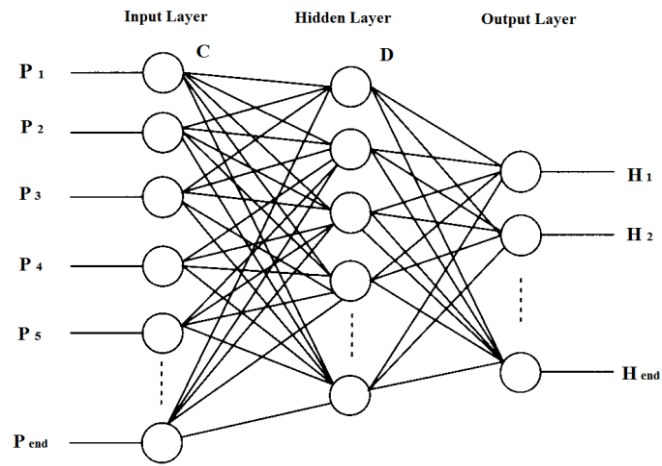

Figure 1. Neural Network Structure.

Let the layer inputs and outputs be $P = (P_0 P_1 P_2 .... P_{512})$, $C = (C_0 C_1 C_3 ..... C_8)$, $D = (D_0 D_1 D_3 ..... D_8)$, $H = (H_0 H_1 H_2 .... H_{32})$, let transfer functions be $f_1, f_2$ and $f_3$, let weight and biases be, $w_1, w_2, w_3, b_1, b_2$ and $b_3$ so neural network can be defines as;

$$H = f_3(w_3 * D + b_3)$$

$$= f_3(w_3 f_2(w_2 C + b_2) + b_3$$

$$= f_3(w_3 f_2(w_2 f_1(w_1 P + b_1) + b_2 + b_3$$

Here $w_1$ is of 8 x 512 size, $w_2$ of 32 x 8 size, $b_2$ of 8x1 size and $b_3$ of 32x1 size, $f_1, f_2$ and $f_3$ are sigmoid functions.

*Block Hash*

Neural network based hash function is depicted in Figure 2.

$$H_r = H_{r1} \otimes H_{r2} \otimes H_{r3} .... H_i \qquad (1)$$

Each dimensions are passed through the block hash and 32x512 sized is performed. XOR values of consecutive rows are calculated in order to obtain 1x512 binary value (1).

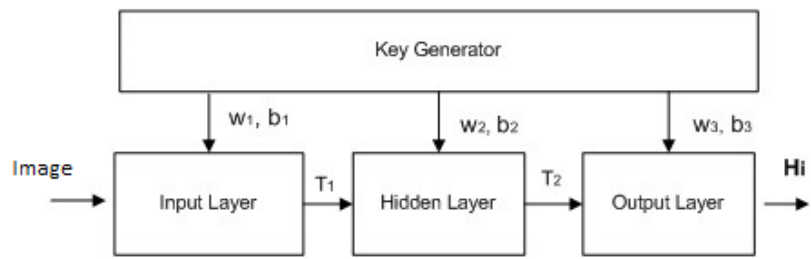

Figure 2. Block Hash.

**Performance Analysis**

In this section, whether suggested hash function validates statistical and security requirements or not is analyzed. So that, statistical distribution, diffusion and confusion, collision resistance and meet-in-the-middle analysis are performed.

*Statistical Distribution of Hash Value*

Security of hash function is directly proportional with uniform distribution of hash value. Figure 3 illustrates 2D graphs of pixel values of the original image localized.

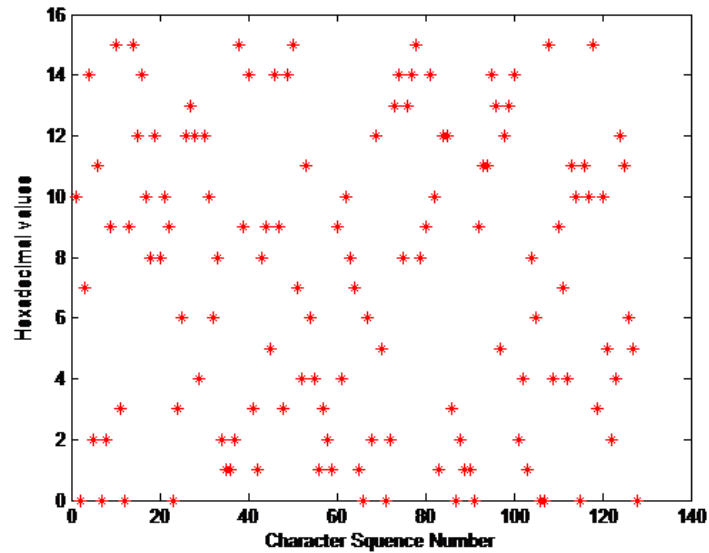

Figure 3. 2D graphs of pixel values of the original image localized.

*Statistical Analyses of Diffusion and Confusion*

Binary format of hash value consists of only 0 and 1 bits, while hexadecimal hash value consists of 16 different characters. Because of this changes in binary hash value must be nearly 50% (as shown in Table 3), in contrast changes in hexadecimal value must be nearly 100%, for each modification. Otherwise diffusion property do not satisfy. In order to control binary and hexadecimal hash value changes following steps are applied: 1. Calculate original image binary and hexadecimal hash value 2. Change value of image 10 pixels randomly. 3. Calculate modified image binary and hexadecimal hash value. 4. Compare and find differences between original and modified images binary and hexadecimal hash values. 5. Repeat 1-4 process Q times. In Figure 4, binary sensitivity of hash value is presented. As it is mentioned binary sensitivity is nearly 50% that satisfies diffusion of hash value. Also, almost 100% hexadecimal sensitivity mean the algorithm has very strong capability of robustness. Statistical parameters for binary sensitivity are defined below: Mean number of bits changed:

$$\overline{B} = \frac{1}{N} \sum_{i=1}^{N} B_i \qquad (2)$$

Mean changed probability:

$$\bar{P} = (\bar{B}/512) \times 100\ \% \qquad (3)$$

Standard deviation of number of bits changed:

$$\Delta B = \sqrt{\frac{1}{N-1}\sum_{1}^{N}(B_i - \bar{B})^2} \qquad (4)$$

Standard deviation of changed probability:

$$\Delta P = \sqrt{\frac{1}{N-1}\sum_{1}^{N}(B_i/128 - \bar{P})^2} \times 100\ \% \qquad (5)$$

Where $N$ = character number of binary hash value ($256$). Through tests $Q = 256, 512, 1024, 2048$ is performed and results are listed in Table 1.

*One Way Property*

Neural Networks most important and intriguing property that makes them useful for applications is their generalization capabilities that is their ability to produce reasonable outputs when they are fed with inputs not previously encountered. If targets' size is so different from inputs' size it is difficult to compute target from input, while it is easy to compute input from target. Due to this property neural network can be used in hash functions Desai2013. Parallel implementation is a significant property of neural networks. Each layer is paralleled. So, they can implement certain functionality independently.

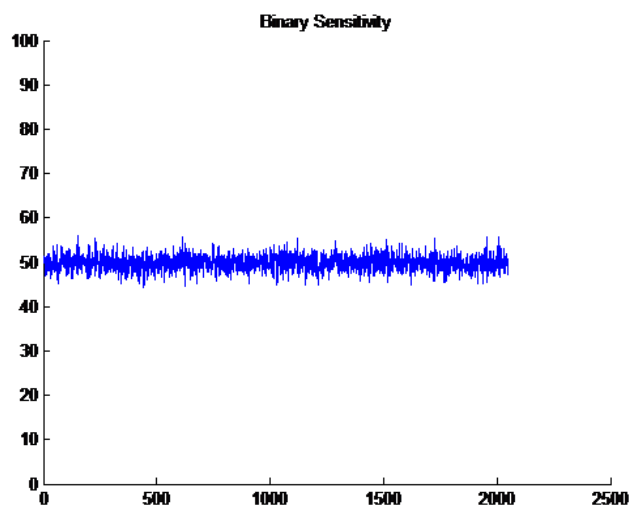

Figure 4. Binary Sensitivity.

According to this, ANN are available for data progressing Neural networks has ability to make relationship using training function with non-linear and complicated values. Confusion is a special property caused by the nonlinear structure of neural networks. This property makes the output depend on the input in a nonlinear and complicated manner. It means that, a bit of output depends on all the bits of the input in a complicated way. Thus, it is difficult to determine the exact input. The confusion property of neural networks makes them a potential choice for cipher designing.

Table 1 Statistics of the number of bit changed.

|  | $Q=256$ | $Q=512$ | $Q=1024$ | $Q=2048$ | Mean |
|---|---|---|---|---|---|
| $B_{min}$ | 210 | 229 | 228 | 227 | 223,50 |
| $B_{max}$ | 280 | 280 | 290 | 286 | 284,00 |
| $\bar{B}$ | 254,6 | 254,87 | 254,87 | 255,21 | 254,89 |
| $\Delta B$ | 8,57 | 8,37 | 8,80 | 8,70 | 8,61 |
| $P_{min}$ | 41,06 | 44,72 | 44,53 | 44,33 | 43,66 |
| $P_{max}$ | 54,68 | 54,68 | 56,64 | 55,85 | 55,46 |
| $\bar{P}\%$ | 49,72 | 49,77 | 49,77 | 49,84 | 49,78 |
| $\Delta P\%$ | 4,92 | 4,92 | 4,92 | 4,93 | 4,92 |

*Analysis of Collision Resistance*

Experimenters must make sure that each bit of original image effects hash value, after generating hash value. In other words hash value must be fully depended on the original image. Single bit change in image do not affects hash value that means vital information security vulnerability. So in this paper, collision resistance analyze is performed Q times as follows: 1. Generate hash value of original image (described in section 3.2) and store in ASCII format. 2. Randomly change least bits in original image 3. Generate hash value of new modified version of original image and store in ASCII format 4. Compare hash values generated in 1 and 3. Find and count same ASCII values at the same location.(3) Plot of distribution of the number of collision hits is illustrated in Figure 5.

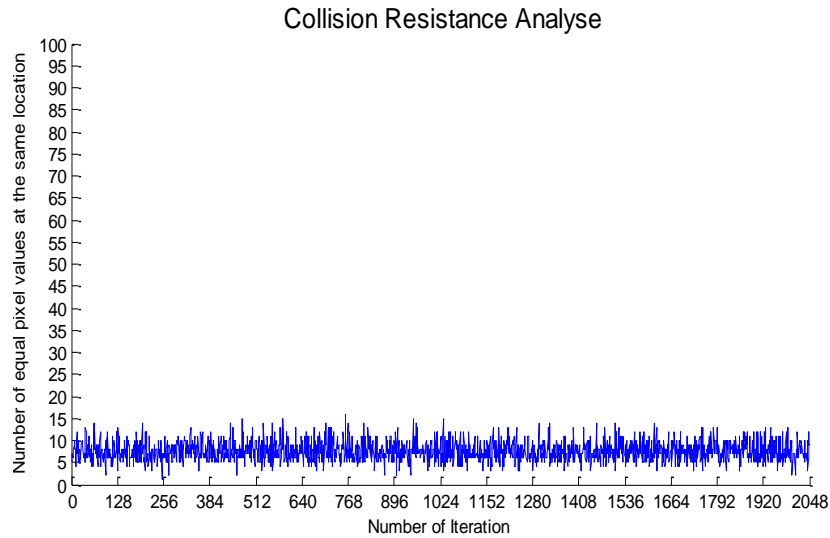

Figure 5. Binary Sensitivity.

Figure 5. Maximum, minimum, mean values and standard deviations are listed in Table.

Maximum, minimum, mean values and standard deviations are listed in Table 2.

Table 2. Collision Resistance Statistical Parameters

| *Iteration* | *Maximum* | *Minimum* | *Mean* | *Standard Error* |
|---|---|---|---|---|
| **256** | 15 | 1 | 7,39 | 2,31 |
| **512** | 16 | 2 | 7,45 | 2,31 |
| **1024** | 15 | 1 | 7,59 | 2,27 |
| **2048** | 16 | 1 | 7,52 | 2,20 |

*Meet-In-The-Middle Attack.*

A meet-in-the middle attack is a technique of cryptanalysis against a block cipher introduced in 1977 (Diffie & Hellman 1977). It is a passive attack; it may allow the attacker to read messages without authorization, but against most cryptosystems it does not allow him to alter messages or send his own (Vanstone et al. 1996). The attacker must be able to calculate possible values of the same intermediate variable (the middle) in two independent ways, starting either from the original or from the hash value. The attacker calculates some possible values each way and compares the results.

If original image is $M = (M_0 M_1 M_2 ..... M_{n-1} M_n)$ its hash value $= H$. Expected image found using meet-in-the-middle attack is $M' = (M_0 M_1 M_2 ..... M_{n-1} M_n')$ its hash value $= H$. In other words attack process is replacing $M_n$ with $M_n'$. But attacker cannot create $M_n'$ that is not in relation with hash value described in previous sections.

**Results and Conclusions**

Secure hash function based machine learning techniques is presented and analyzed here. Proposed algorithm is efficient to require diffusion and confusion properties due to neural network information transfer process inspired from real biological systems. Analyses and experiments explained in this paper reveals that hash function satisfies sensitivity, minimum collision hit requirements and powerful against attacks like meet-in-the-middle.

Figure 3, uniform distribution of hexadecimal hash value against local distribution of original image means high randomness that requires confusion. Figure 4, nearly 50% difference of binary format of hash value means high sensitivity that requires diffusion. But Figure 4 is not sufficient only by itself. In order to correlate Figure 4, statistically approaches are shown in Table 1. When looping sensitivity testing process as Q times, average 254, minimum 224, maximum 284 bits of 512 bits differs with minor standard deviation (8.61) and minimum 43.66%, maximum 55.46%, average 49.78% of 512 bits differs with minor standard deviation (4.92). These results satisfy sensitivity of neural network based hash function.

Calculation of same ASCII hash values at the same location that is called collision resistance is performed as Q times. Figure 5. illustrates collision resistance when Q = 2048. When looping collision resistance testing process as Q times, average 7 bit same ASCII values are found at the same location that can be ignored due to minority. So these results satisfy collision resistance of neural network based hash function. As a result; this system can be used in communication applications especially in military applications.